\documentclass[journal=jpcafh,manuscript=article, layout=twocolumn]{achemso}

\newcommand{\setname}[1]{\mbox{#1}}

\newcommand\myfig[2][width=\columnwidth]{%
    \includegraphics[#1]{#2}%
}
\usepackage{color}

\author{Mads~Engelund}
\email{mads.engelund@gmail.com}
\phone{+34632702734}
\affiliation[Centro de F\'{\i}sica de Materiales (CFM)]{Centro de F\'{\i}sica de Materiales (CFM) CSIC-UPV/EHU,  Paseo Manuel de Lardizabal 5, E-20018, Donostia-San~Sebasti\'an, Spain}
\author{Nick~Papior}
\affiliation[DTU Nanotech]{DTU Nanotech, \O rsted Plads, building~345E, DK-2800, Kgs.~Lyngby, Denmark}
\alsoaffiliation[Center for Nanostructured Graphene (CNG)]{Center for Nanostructured Graphene, CNG, DK-2800, Kgs.~Lyngby, Denmark}
\author{Pedro Brandimarte}
\affiliation[Centro de F\'{\i}sica de Materiales (CFM)]{Centro de F\'{\i}sica de Materiales (CFM) CSIC-UPV/EHU,  Paseo Manuel de Lardizabal 5, E-20018, Donostia-San~Sebasti\'an, Spain}
\author{Thomas~Frederiksen}
\affiliation[Donostia International Physics Center (DIPC)]{Donostia International Physics Center, DIPC, Paseo Manuel de Lardizabal 4, E-20018, 
Donostia-San~Sebasti\'an, Spain}
\alsoaffiliation[Ikerbasque]{IKERBASQUE, Basque Foundation for Science, E-48013, Bilbao, Spain}
\author{Aran~Garcia-Lekue}
\affiliation[DIPC]{Donostia International Physics Center, DIPC, Paseo Manuel de Lardizabal 4, E-20018,  Donostia-San~Sebasti\'an, Spain}
\alsoaffiliation[Ikerbasque]{IKERBASQUE, Basque Foundation for Science, E-48013, Bilbao, Spain}
\author{Daniel~S\'anchez-Portal}
\affiliation[Centro de F\'{\i}sica de Materiales (CFM)]{Centro de F\'{\i}sica de Materiales (CFM) CSIC-UPV/EHU,  Paseo Manuel de Lardizabal 5, E-20018, Donostia-San~Sebasti\'an, Spain}
\alsoaffiliation[DIPC]{Donostia International Physics Center, DIPC, Paseo Manuel de Lardizabal 4, E-20018,
Donostia-San~Sebasti\'an, Spain}

\title{Search for a Metallic Dangling-Bond Wire on $n$-doped H-passivated  Semiconductor Surfaces}
\abbreviations{DFT, GGA, LDA, DB}

\begin{document}

\begin{abstract}
We have theoretically investigated the electronic properties of neutral and $n$-doped dangling bond (DB) quasi-one-dimensional structures (lines) in the Si(001):H and Ge(001):H substrates with the aim of identifying atomic-scale interconnects exhibiting metallic conduction for use in on-surface circuitry.
Whether neutral or doped, DB lines are prone to suffer geometrical distortions or have magnetic ground-states that render them semiconducting.
However, from our study we have identified one exception -- a dimer row fully stripped of hydrogen passivation. 
Such  a DB-dimer line shows an electronic band structure which is remarkably insensitive to the doping level and,  thus, it is possible to manipulate the position of the Fermi level, moving it away from the gap. 
Transport calculations demonstrate that the metallic conduction in the DB-dimer line can survive thermally induced disorder, but is more sensitive to imperfect patterning. 
In conclusion, the DB-dimer line shows remarkable stability to doping and could serve as a one-dimensional metallic conductor on $n$-doped samples.
\end{abstract}

\section{Introduction}
The dream of fabricating structures with atomic precision  has become a reality  in certain semiconductor substrates.
Local hydrogen desorption, induced with the aid of a scanning tunnelling microscope (STM) tip, on the hydrogen-passivated (1$\times$2)-reconstruction of the Si(001) and Ge(001) substrates [hereafter Si/Ge(001):H] can be used to  reproducibly and rapidly  create dangling-bond (DB) arrays defined with atomic precision.\cite{Kolmer2012Electronic,  Kolmer2013Construction, Schofield2013Quantum, Engelund2015Tunneling} 
This is a powerful method that allows one to directly pattern  planar electronic devices and circuitry made out of DBs,~\cite{Haider2009Controlled, Godlewski2013Contacting, Pitters2011Charge, Livadaru2011Theory, ShaterzadehYazdi2014Characterizing, Kawai2012Danglingbond} 
and to create templates for anchoring molecular networks~\cite{Lopinski2000Selfdirected, Kruse2002, Tong2004, Guisinger2004,Godlewski2016Interaction}  or for selective dopant incorporation~\cite{Radny2007Single,Reusch2007Single,Scappucci2011Complete, Weber2012Ohms}.  
DB arrays have  a very rich and complex behaviour, exhibiting several  metastable charge- and spin- states,~\cite{Kepenekian2013SurfaceState, Kepenekian2013Electron, Robles2012Energetics, Kepenekian2014Spin, Lee2009Quantum, Raza2007Theoretical, Bellec2013Reversible, Engelund2015Tunneling,Kawai2016Electronic} as well as small polaron and soliton formation.\cite{Bowler2000Small, Bird2003Soliton} 
While this complexity makes the study of these systems  attractive from a fundamental point of view, it might hinder their use for technological applications. 
In particular, it would be desirable to have a practical scheme to fabricate simple conductive wiring for interconnecting  complex functional units of atomic dimensions  (e.g., atomic and molecular logic gates~\cite{Joachim2012Different, Atmolbook13, Dridi2015Mathematics, Kleshchonok2015Quantum}) deposited or patterned on semiconductor substrates.~\cite{Atmolbook12}  

There have been several theoretical proposals of logic gates constructed  with DBs on Si/Ge(001):H surfaces.~\cite{Kawai2012Danglingbond, Kleshchonok2015Quantum}
A recent joint experimental and theoretical work has demonstrated the operation of one such device.~\cite{Kolmer2015Realization}  
In that experiment, however, the DB gate was addressed using STM tips, which does not represent a scalable technology. 
For this to be realized in practice, it would be highly desirable that atomic-scale wiring could be created directly on the surface via a simple procedure. 
Given the existing techniques, the possibility to fabricate
atomic-scale metallic connectors made out of DBs on Si/Ge(001):H would be highly attractive. 
This,  together with the ability to  fabricate logic elements using DBs as building blocks, would allow creating a full circuit in a single processing step  with atomic-scale precision. 
Unfortunately, quasi-one-dimensional (1D) structures formed by  DBs are prone to suffer from instabilities that open gaps in the electronic band structure, making them  less than ideal candidates for atomic-scale interconnects.~\cite{Kolmer2012Electronic, Kepenekian2013SurfaceState, Naydenov2013Engineering, Mantega2012Spectroscopic}
In such cases, electron transport can still occur through diffusion of polarons,\cite{Bowler2000Small} but transport in this regime is typically less efficient.

Half-filled electron bands in 1D structures are particularly prone to suffer distortions that destroy the metallicity of the system.~\cite{Peierls55, Mott68, Kennedy87, Gruener88} 
A possible strategy to enforce a metallic character of the DB lines could be to dope the system with holes or electrons, moving the systems away from the half-filled condition.
The idea is that this would move the Fermi energy away from the band gap provided that the band structure remains more or less constant.

The concept of moving the Fermi level away from  the band gap by doping the system is particularly attractive due to its simplicity.  
However,  it is in general far from certain that the geometric and electronic structure of the DB lines remain unchanged under different electron fillings. 
The purpose of the present theoretical study is to clarify this issue for several key structures of DB lines on both Si(001):H and Ge(001):H substrates.
Few studies have so far addressed this interplay between doping and stability, and then only for a one type of DB line.\cite{Bowler2000Small, Kepenekian2014Spin}
Yet doping is both an important experimental parameter and largely unavoidable, even in  supposedly pristine substrates\cite{Wojtaszek2014Inversion}.

We limit our investigation to $n$-type doping because unoccupied DB derived electronic features tend to lie in  the  bandgap whereas the occupied ones are either resonant with, or close to, the bulk valence bands of both Si\cite{Engelund2015Tunneling,Naydenov2013Engineering} and Ge\cite{Kolmer2012Electronic}. 
Therefore we judge $p$-type doping as less likely to yield functional wires since DB derived bands would only be weakly localized to the DB lines.

\begin{figure}
  \centering \myfig{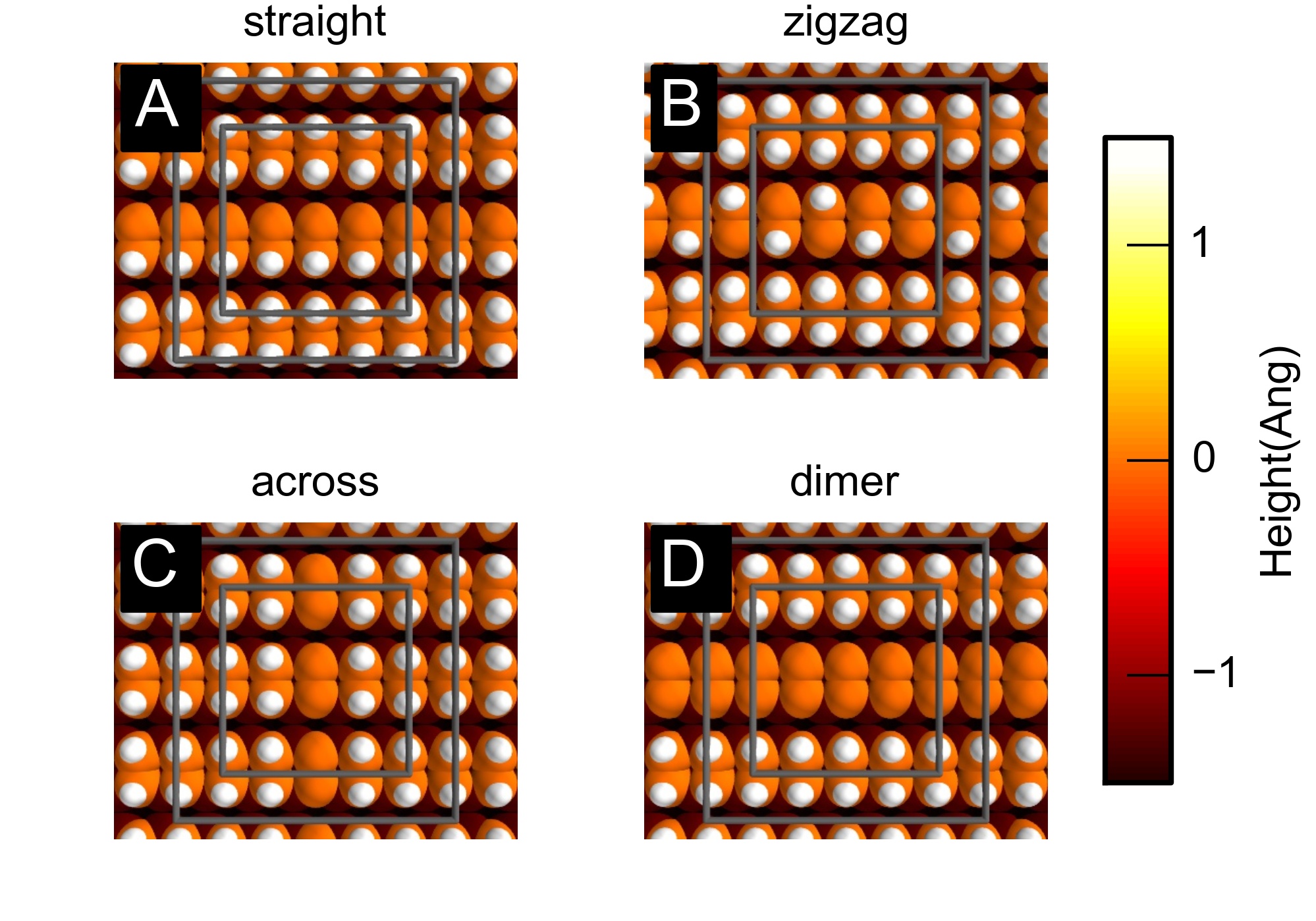}
  \caption{Top view of the four basic DB line structures investigated in this paper. The structures are idealized, i.e., they correspond to the situation in which all Si/Ge atoms along the DB lines are equivalent. 
 Small spheres represent hydrogen atoms while large ones represent Si or Ge atoms. Atoms are colored according to their height on the surface and the surface unit cells considered in this study (($2\times4$) or ($3\times6$)) are shown as grey squares.
      \label{fig:structures}
  }
\end{figure}

We have investigated four basic DB lines. 
Three of these lines follow the basic dimer rows of the Si/Ge(001):H surfaces  and one is perpendicular to these rows (see Fig.~\ref{fig:structures}).
The investigated structures are, 
(A)  the removal of all H atoms from one side of the dimers along one row (``\setname{Si/Ge-straight}''), 
(B) the removal of H on alternating sides of the dimers along one row (``\setname{Si/Ge-zigzag}''), 
(C) the perpendicular structure formed by the removal of both H atoms
from one dimer on each dimer row  (``\setname{Si/Ge-across}'') and finally, 
(D) the complete hydrogen removal from one dimer row (``\setname{Si/Ge-dimer}'').

\section{Computational Methods}
The DB lines have been studied by spin-polarized density functional theory (DFT) using the SIESTA code~\cite{Soler2002SIESTA}. 
The slabs used in the calculations contain 8 Si/Ge layers, including the reconstructed Si/Ge surface layer. A (2$\times$4) surface super-cell was used.  
Atoms in the lower layer of the slab, which follow a bulk-like arrangement, are passivated with hydrogen.
Relaxations were performed with a force tolerance of 0.02 eV/{\AA}, fixing the lower 4 Si/Ge layers  as well as the hydrogen passivated layer at the bottom of the slab.  
The configurations were initialized using the Si/Ge(001):H positions, with a fully polarized spin structure and one DB site was raised either 0.01 or 0.4~\AA.
Those configurations which relaxed to a spin-polarized final state were re-initialized to ensure that both ferromagnetic and anti-ferromagnetic ordering was attempted.

We used the generalized gradient approximation (GGA) to describe exchange and correlation~\cite{Perdew1996Generalized}, a mesh-cutoff of 200~Ry for real-space integrations, 
a double-$\zeta$ plus polarization (DZP) basis set generated with an energy shift~\cite{Soler2002SIESTA} of 150~meV, and a (6$\times$3) $k$-point grid, with the highest sampling along the direction of the 1D DB line.
Local-density approximation (LDA) calculations were performed to test stability of our conclusions with respect to the exchange-correlation functional used.
For the Si substrate we performed GGA functional calculations with a larger (3$\times$6) supercell and 10 layers (8 of them relaxed) were performed for all lines to verify  that our conclusions are robust against the number of layers present and relaxed in the slab,  the distance between the DB lines, and the length of the  DB-line portion explicitly considered. 
The level of approximation was chosen to reproduce structural deformations as well as qualitative features of the electronic structure. 

\section{Results and Discussion}
The main question to solve is whether it is possible to achieve a conducting DB wire by doping or if the excess carriers will tend to localize, thereby creating a new distortion pattern. 

To simulate the $n$-doping of the DB lines, we doped the system with extra electrons and  compensated their charge by adding a homogeneous positive charge background so the system remained globally neutral. 
We decided to use integer number of electrons in our calculations since small fractional charging of the unitcell would artificially suppress the formation of localized charge distortions.
On the other hand, this means that we limit ourselves to simulating relatively large doping levels.
These calculations allow us to determine if excess electrons tend to enter delocalized states of the extended 1D line or cause polaron formation. 

Our calculations model uniform doping- i.e., a situation  where dopants do not tend to segregate to the surface and, thus,  their number close to the surface might be considered as negligible.
If special care is taken to avoid surface depletion of dopants under preparation, the specifics of dopant type and placement could become important  to determine the properties of DB-lines under doping.~\cite{Labidi2015Scanning,Kepenekian2014Spin}

Since few qualitative differences were seen between Si(001):H and Ge(001):H, we show only the Si results in  the figures and mention the differences between both substrates in the text.

\begin{figure}[t!]
  \centering
  \myfig{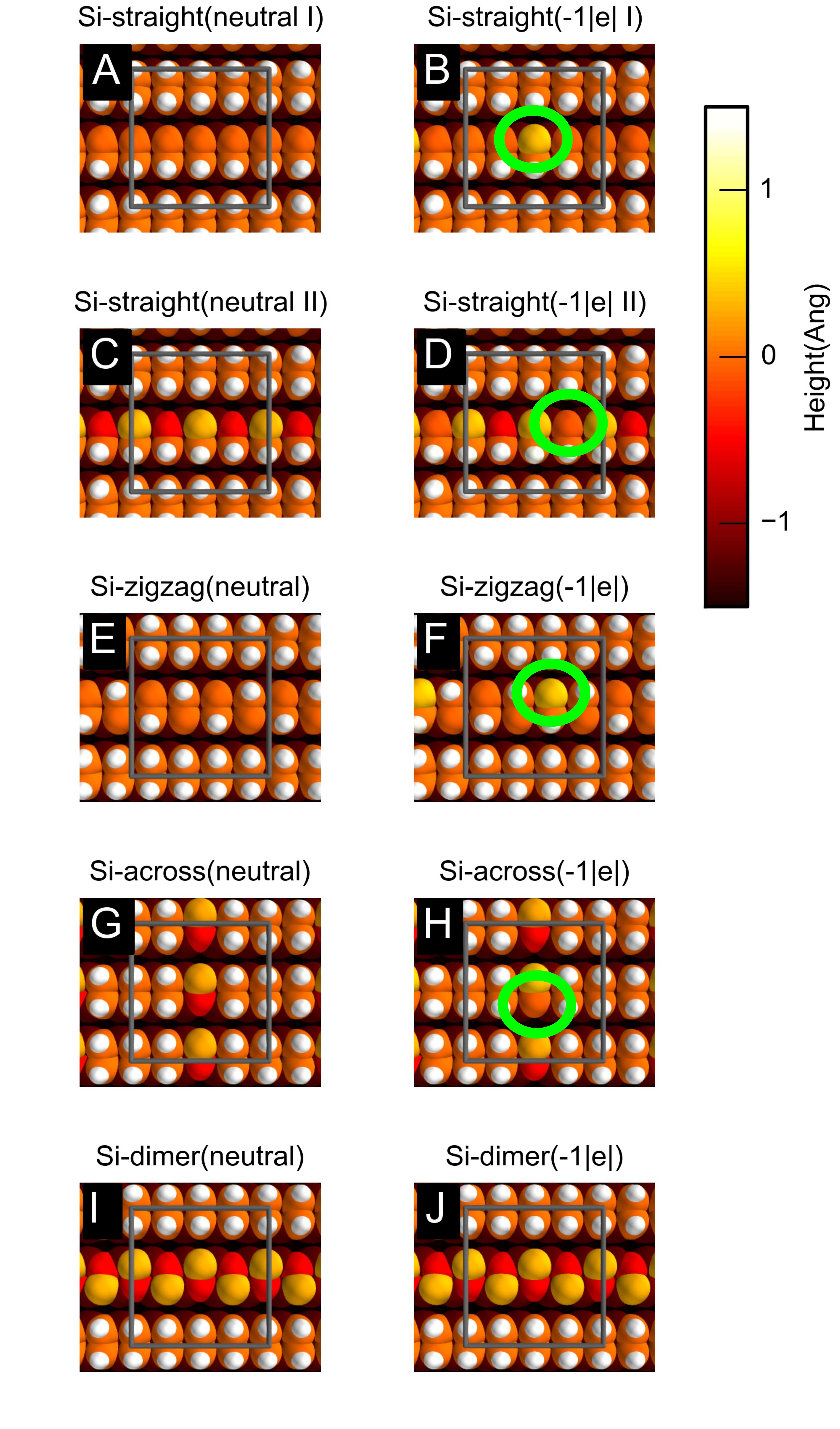}
  \caption{
  Relaxed structures of different DB lines in Si(001):H. 
  Neutral structures and those with one extra electron in the simulation cell are compared. 
  Two structures are shown for the \setname{Si-straight} line as explained in the text,  neutral I corresponds to the antiferromagnetic ordered ground-state, while neutral II shows the structurally distorted configuration lying only 5~meV per DB higher in energy.
  Atoms are colored according their height. 
  The green rings in panels B, D, F and H highlight the DB site which undergoes a geometric distortion upon doping.
    \label{fig:relaxed_structures}
  }
\end{figure}

As a starting point we naturally investigated the properties of the undoped DB lines (see Fig.\ref{fig:relaxed_structures}, left column).
The first thing to note is that all the DB lines exhibit band gaps with no states at the Fermi level in the absence of doping (hereafter we refer to this undoped  situation as ``neutral''). 
These gaps, however, have two types of origins:
({\it i}) a geometric distortion combined with charge transfer between DBs occupying inequivalent positions, and 
({\it ii}) local spin-polarization which opens a gap in the spin-polarized band structure.  
These two effects compete since charge reorganization tends to create doubly- and un-occupied DB sites, while spin-polarization tends to create singly-occupied DB sites.  
From a structural point of view, DB sites come in roughly three groups, raised position (doubly-occupied), neutral position (singly-occupied) and lowered position (unoccupied). 

In agreement with earlier  studies~\cite{Kolmer2012Electronic, Kepenekian2013SurfaceState}, we find that the most stable solution for \setname{Si/Ge-across} and \setname{Si/Ge-dimer} lines  corresponds to a strong geometric distortion,  while the \setname{Si/Ge-zigzag} structures undergo a magnetic distortion (in which all DB sites have the same population and, thus, the same height over the substrate).
For the \setname{Si/Ge-straight} line both magnetically distorted states and unpolarized states with geometric distortion could be stabilized.
For the \setname{Ge-straight} line we calculate that the geometrically distorted state is more stable by 100~meV per DB than a magnetically distorted state with anti-ferromagnetic ordering while the corresponding value for the \setname{Si-straight} line is less than 5~meV per DB.
In the latter case, this energy difference is too small to confidently assign either of these solutions as the ground-state of the system due to different limitations in our model, such as the spin contamination of unrestricted DFT calculations\cite{Koch2001Chemist,Cohen2007Evaluation,Huzak2011Halfmetallicity}.

In contrast, the two lines made up of unpassivated dimers, the \setname{Si/Ge-across} and \setname{Si/Ge-dimer} lines, are less sensitive to initial conditions and invariably relax to a geometrically distorted configuration where each dimer contains both a doubly- and un-occupied sites. 
It is, however, possible to stabilize different patterns due to the relative orientations of the dimers in the lines - something we will discuss in detail later.
In general, different structures on Ge(001):H tend to be better separated in energy than those on Si(001):H.
For example, the \setname{Ge-straight} DB line is more stable by 140~meV per DB site than the \setname{Ge-zigzag} DB line, while the corresponding number for Si(001):H is only 50~meV per DB.

Moving on to the $n$-doped results (see the right column of Fig.~\ref{fig:relaxed_structures}), upon charging with one electron the   \setname{Si/Ge-straight}, \setname{Si/Ge-zigzag} and \setname{Si/Ge-across} lines  show a strong geometric distortion for at least one DB site (compare Figs.~\ref{fig:relaxed_structures}A,C,E, G to Figs.~\ref{fig:relaxed_structures}B,D,F,H), while the \setname{Si/Ge-dimer} lines do not exhibit such distortion (no visible change from Fig.~\ref{fig:relaxed_structures}I to Fig.~\ref{fig:relaxed_structures}J). 

\begin{figure}[!t]
  \centering
  \myfig{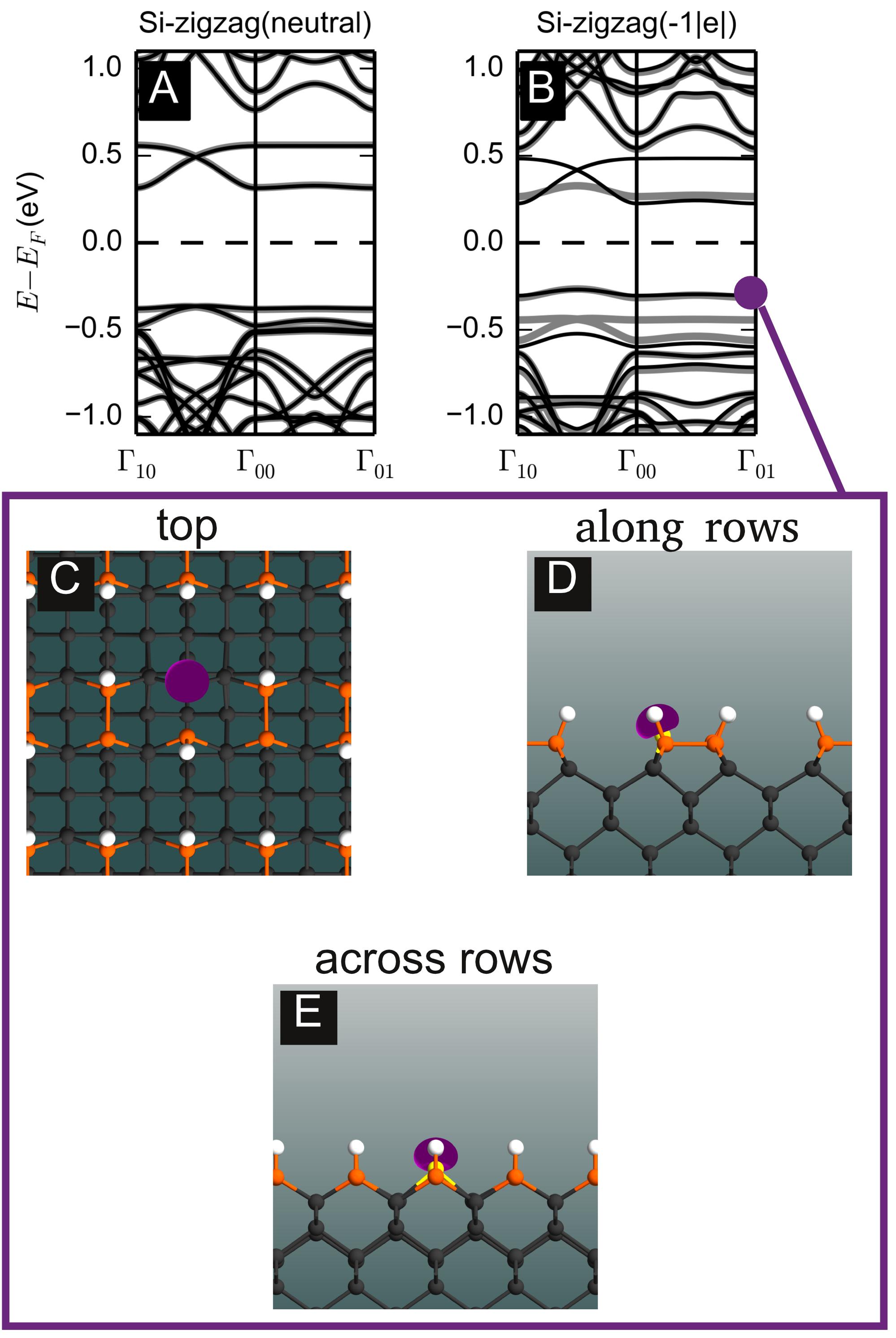}
  \caption{Band structures of the \setname{Si-zigzag} 1D line: (A) neutral and (B) doped with one extra electron per cell. 
    The band structure path goes from $\Gamma$-point to $\Gamma$-point, first along the DB line ($\Gamma_{10}-\Gamma_{00}$), then along  the transverse direction ($\Gamma_{00}-\Gamma_{01}$). 
Majority (minority) bands are shown as thin black (thick grey) lines.  
  (C-E) Isosurface plots (0.05~e\AA$^{-3}$ isovalue) of the density associated with the spin-degenerate band marked in panel B. 
  }
 \label{fig:localized_electron}
\end{figure}

The \setname{Si/Ge-zigzag} DB line can be used as an example of a structure that can undergo geometric distortions.
Fig.~\ref{fig:localized_electron} shows that, as charge is added, local geometrical distortions are coupled to the localization of the extra carriers at specific sites.   
The \setname{Si/Ge-zigzag} DB line gains  one electron and one DB site becomes fully occupied, removing the spin-polarization at that site and moving the corresponding atom to a higher position (as highlighted in Fig.~\ref{fig:relaxed_structures} F). 
As a consequence,  the Fermi level lies within the new gap and the band structure remains insulating. 
As previously observed by \citeauthor{Bowler2000Small}~\cite{Bowler2000Small} for the \setname{Si-straight} DB line, the  extra electron becomes self-trapped and forms a small polaron. 
We note that this is only one possible configuration of the charged \setname{Si/Ge-zigzag} and \setname{Si/Ge-straight} lines since the competition of between charge and spin-polarization produces a large number of meta-stable states.
However, it is clear that the neutral ground state is not stable under charging.

\begin{figure}[!t]
  \centering
  \myfig{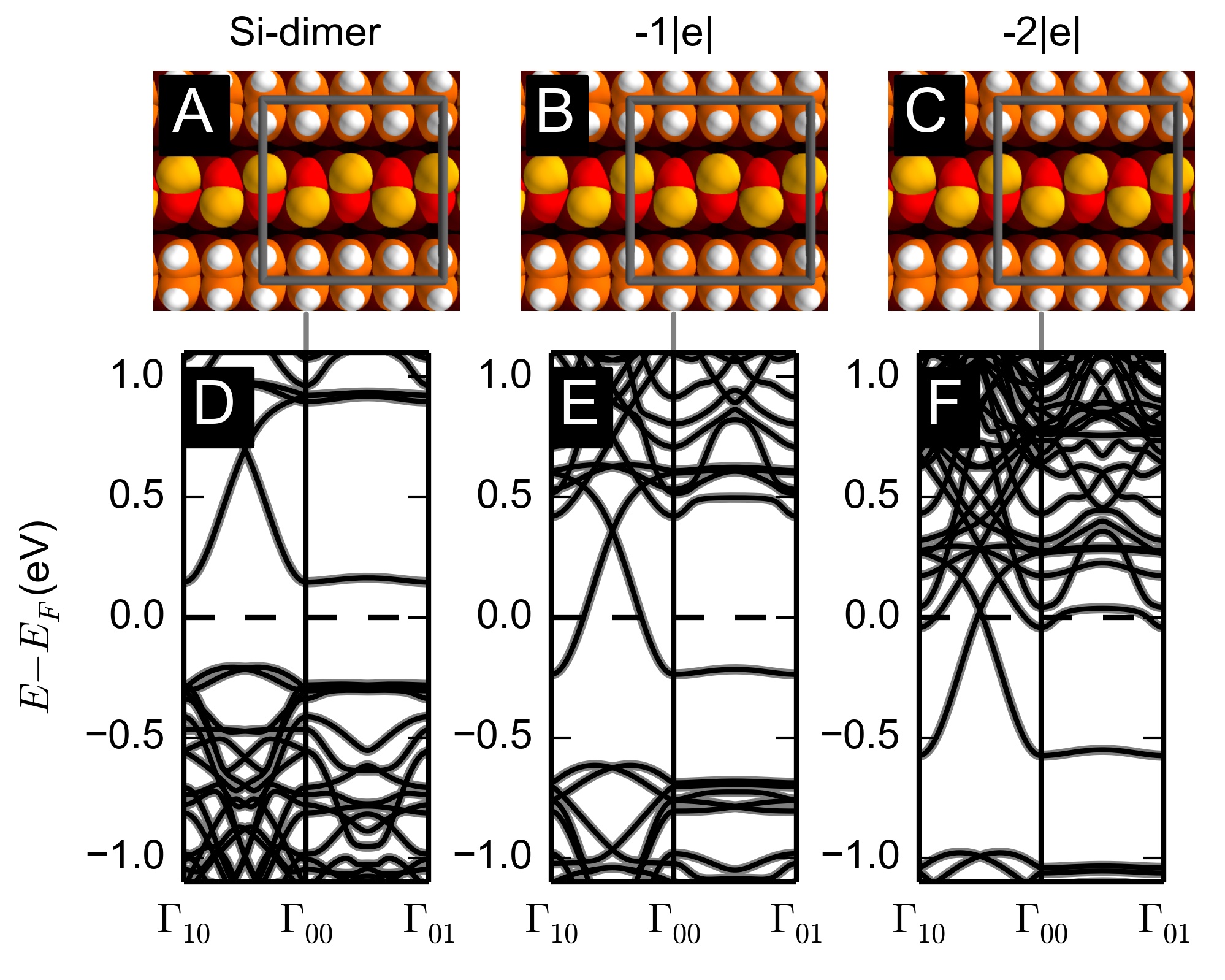}
  \caption{
      (A-C) Relaxed geometries and (D-F) electronic band structures of the \setname{Si-dimer} DB line in the neutral state and charged with one and two  extra electrons in the (2$\times$4) supercell (corresponding to ${1\over8}$ and ${1\over4}$ electrons  per DB site, respectively).
      These  spin-degenerate band structures follow the same path as those in  Fig.~\protect\ref{fig:localized_electron}.
      }
  \label{fig:dimer_band}
\end{figure}

In striking contrast with the rest of the DB lines, the simple picture of doping holds for the \setname{Si/Ge-dimer} system: 
as seen in Fig.~\ref{fig:dimer_band}, when electrons are added to the cell, both  the geometric and electronic structures remain stable.
The changes of the geometric structure are small since the additional charge  distributes homogeneously along the DB line and, as a consequence, the  changes of the band structure correspond mostly to a rigid shift of the Fermi level.  
In general, we observe that the basic features of the electronic band structure  persist until the the Fermi level reaches the conduction band edge, at which point any additional electrons will occupy the conduction band (see Fig.~\ref{fig:dimer_band}F). 
If we assume that this stability  occurs also at the intermediate doping values  between calculations, we conclude that the structure should be stable and metallic at any level of excess electrons near the surface. 
However, we should always be aware that the presence of excess electrons at the surface is not guaranteed even on nominally $n$-doped substrates.\cite{Wojtaszek2014Inversion}
These results are consistent with the experimental scanning tunnelling spectroscopy  data by \citeauthor{Naydenov2013Engineering},~\cite{Naydenov2013Engineering} who demonstrated that the \setname{Si-straight} structures exhibit no states at the Fermi level while the \setname{Si-dimer}  do.

In general, very similar results are found for the two substrates.
The important edge case was the \setname{Si/Ge-straight} DB lines where we found two possible ground-state structures for the \setname{Si}-straight line.
However, by explicitly initializing the charged calculations for both substrates from both possible ground-states (non-magnetic and buckled,  spin-polarized and flat), in all cases we observe stabilization of a self-trapped polaron when adding extra-electrons to the system.
The \setname{Ge-straight} DB line is also the only structure where we observed a qualitative difference when using different exchange-correlation functionals. 
In contrast to the GGA case, our LDA results predict the buckled \setname{Ge-straight} DB line to be stable with respect to  charging. 
Otherwise, our doping results are robust with respect to both substrate and functional. 

After our investigation, the \setname{Si/Ge-dimer} DB line clearly stands out. 
When doped with electrons it retains its geometric and electronic structure in  marked contrast to the three other DB lines investigated. 
Having identified the $n$-doped \setname{Si/Ge-dimer} DB lines as possible candidates for conducting wires,  
we now proceed to investigate their transport properties and, in particular,  how these properties are influenced by simple defects which are likely to be present in such wires.
 This is a crucial piece of information for the applicability of these structures as interconnects in atomic-scale circuits. 
 
\begin{figure}[!t]
  \centering
  \myfig{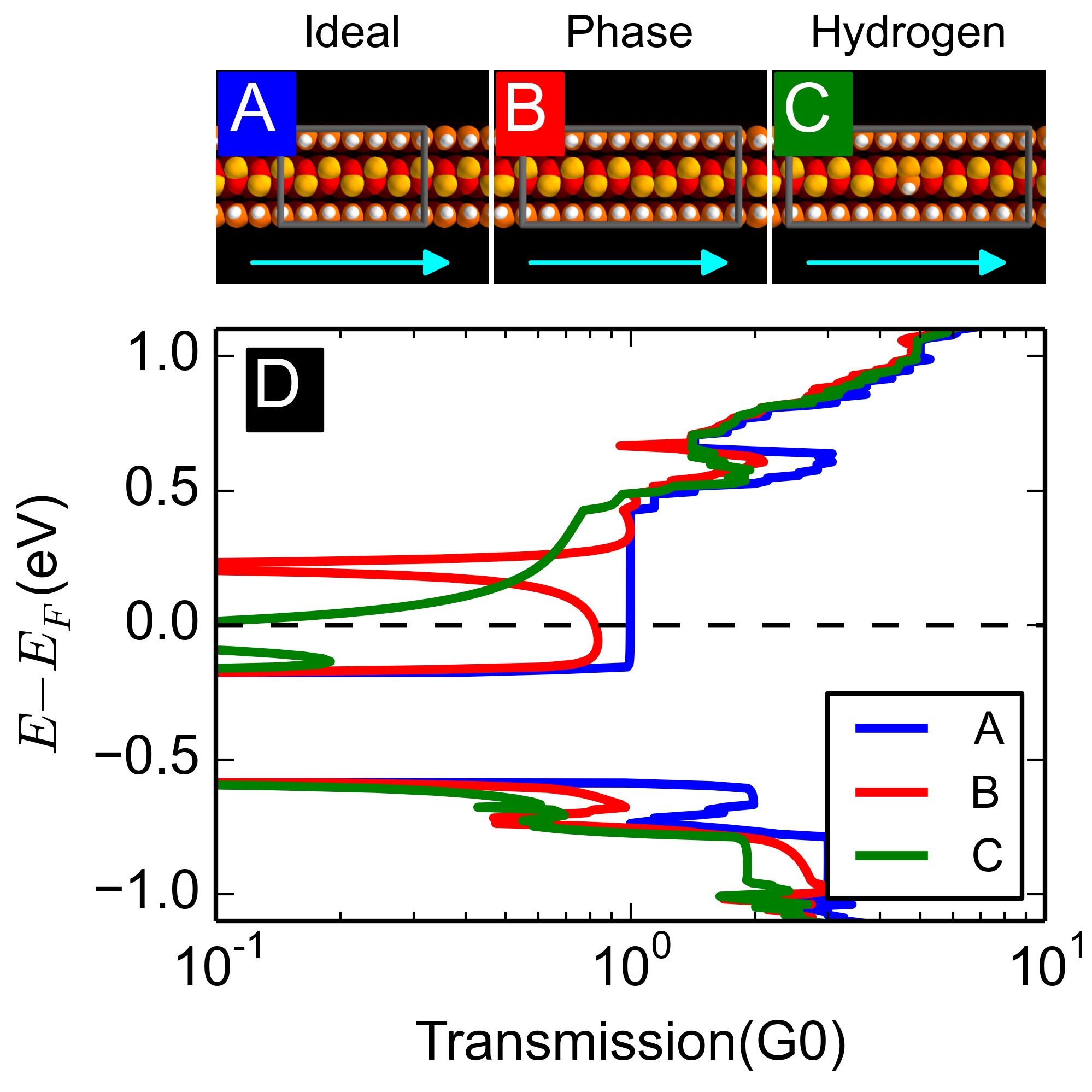}
  \caption{ 
      (A-C) Relaxed structures and (D) transmission spectrum relative to the Fermi level for a DB-dimer line with different types of defects on Si(001):H.
      Transmission curves are shown for a (A) an ideal Si-dimer wire, (B) a phase-shift defect, and (C) a hydrogen defect with a  doping corresponding to $\frac18$ electrons per DB site.
      None of these structures exhibits spin polarization at this doping level, so both spin channels have the same transmission. 
      \label{fig:phase_shift_trans}
  }
\end{figure}

To examine possible disruptions of the \setname{Si/Ge-dimer} wire, we have investigated two types of defects, a ``hydrogen defect'' and a  ``phase-shift'' defect (see Fig.~\ref{fig:phase_shift_trans}).
The hydrogen defect is simply the re-passivation of a single DB site in the structure  by a hydrogen atom -- a type of defect which could occur due to imperfect hydrogen removal. 
The phase-shift defect consists of two dimers which are slanted in the \emph{same} direction, while the remaining dimers are slanted following the normal alternating pattern. The phase-shift defect is effectively the boundary point between two domains of the DB-dimer line with oppositely alternating patterns.  
From entropic considerations,  a sufficiently long DB-dimer line will have multiple alternating domains and thereby phase shift defects -- the boundary between the domains. 
We have estimated the energy cost of creating a phase-shift defect to be $\sim 100$~meV for both Si and Ge by comparing the total energies of an ideal DB-dimer line and a DB-dimer line containing a phase-shift  defect in a  $(2\times 9)$ supercell. 
This energetic cost means that  we can expect approximately one phase-shift defect in every 50 dimers in thermal equilibrium at room temperature. 
At liquid nitrogen temperatures the occurrence of these defects from thermal fluctuations should be negligible, but we assume that they could be easily created during the surface patterning process - on purpose or inadvertently-, e.g., if domains of different alternation are created separately and then connected.

Electron transport across these defects was investigated  using the  TranSIESTA code~\cite{Brandbyge2002Densityfunctional}.
These calculations were performed using open-boundary conditions, and ($2\times 6$), ($2\times 9$) and ($2\times 10$) simulation cells  for the ideal wire, phase-shift and hydrogen defects, respectively.
These supercells include the central  scattering region and ($2\times2$) blocks in either end, corresponding to the electrodes, and a consistent level of doping per dimer row was used in the electrode and full calculations.
For transmission calculations a ($7\times1$) $k$-point sampling in the  directions transverse to the transport direction was used.  

Figure~\ref{fig:phase_shift_trans} shows  the transmission curves for each type of defect and a  doping level corresponding to $\frac18$ electrons per DB  (or $\frac14$ electrons per unsaturated DB-dimer) in Si(001):H.
 As earlier, the results are very similar in the case of Ge DB-dimer lines. 
For this level of doping, a single phase-shift defect reduces the transmission at the  Fermi level from 1 to 0.8 (0.9) for Si (Ge).     
In other words, $\sim10\%$  of the transmission survives after propagation through 10 (21) phase-shift defects. 
At room temperature and assuming that this is the only type of defect present, this would correspond to wires containing more than $\sim500$ (1000) DB-dimers. 
The geometric and electronic structure of the phase-shift defect is insensitive to the doping level - similarly to the Si/Ge-dimer wire itself. Due to this and the relatively constant transmission curve our estimate should hold also at lower doping levels.
To explore the stability of the structure under bias, we have performed a relaxation with a 0.2~V voltage drop across the defect structure. We find that essentially no further relaxation takes place and that the transmission is relatively constant in the bias window. 
All in all, the presence of the phase-shift defect does not seem to greatly diminish the metallic character of the wire.  

The hydrogen defect is  more complex  since the structure and spin-polarization  changes with the doping level.  
The DB site next to the hydrogen defect (within the same DB-dimer) can be neutral or negatively charged depending on the doping level and the substrate. 
At the (high) doping level presented in  Fig.~\ref{fig:phase_shift_trans}, the hydrogen defect is negatively charged (and non-magnetic) and the presence of the defect almost completely turns off the transmission at the Fermi level.
Lowering the doping level results in a neutral defect on both substrates, i.e., the DB next to the hydrogen atom is only occupied by an un-paired electron and,  as a consequence, it is spin-polarized. 
The neutral  hydrogen defect results in different scattering for carriers with different spin orientations. 
Despite this complex behaviour though, the conclusion is simple -- the uncontrolled inclusion of a single hydrogen defect is likely to break the intended functionality.

\section{Conclusion}
As we have seen, surface DB structures are highly complex due to the  delicate interplay between their geometric, electronic and magnetic properties.  
Electrons tend to localize at DB sites generating both  geometrical distortions and ordered magnetic arrangements that usually  translate onto the opening of gaps at the Fermi level in the electronic band structure.
This makes it difficult to achieve a simple metallic wire made exclusively of surface DBs. 
Although this does not diminish the potential of DB nanostructures  to be used for other atomic-scale electronic devices which could harness the complexity,  for example to fabricate logic gates or tunneling devices, this situation certainly limits their applicability as stable metallic interconnects. 

However, we have found that stripping a full dimer row of hydrogen passivation is a promising route to achieve metallic interconnects on $n$-doped Si(001):H and Ge(001):H substrates at low temperatures. 
Our calculations predict that excess electrons do not localize along the DB-dimer lines  and that the scattering from likely structural distortions would be relatively low.  
On the other hand, some impurities, such as re-passivated DB sites, would need to be tightly controlled for the DB-dimer to retain conductivity.
If structural impurities can be eliminated, using such interconnects would allow reaping the benefits of a single processing step when making on-surface circuits based on DB nanostructures. 

From a practical point of view, DB arrays based on DB-dimers are also easier to fabricate since the ability to  create DBs one-by-one has only been demonstrated on the silicon substrate~\cite{Lyding1994Nanoscale, Haider2009Controlled}, while for germanium only dimer-by-dimer desorption  has so far been demonstrated.~\cite{Kolmer2012Electronic}
Achieving $n$-doping on Ge(001):H might be a practical problem, since experiments have shown that an inversion layer occurs at the surface of $n$-doped samples.~\cite{Wojtaszek2014Inversion} 
However, if the construction of DB devices is combined with a technique to precisely incorporate phosphorous dopants in the sub-surface region,\cite{Schofield2003Atomically, Radny2007Single} this problem might be overcome.

\begin{acknowledgement}
This work is funded by the FP7 FET-ICT ``Planar Atomic and Molecular Scale devices'' (PAMS) project (funded by the European Commission under contract No.~610446). 
ME, PB, TF, AGL and DSP also acknowledge support from the Spanish Ministerio de Econom\'{\i}a y Competitividad (MINECO) (Grant No.~MAT2013-46593-C6-2-P) and the Basque Dep. de Educaci\'on and the UPV/EHU (Grant No.~IT-756-13). 
MayaVi\cite{Ramachandran2011Mayavi}, Matplotlib\cite{Hunter2007Matplotlib} and Virtual NanoLab\cite{QuantumWise2014} was used in the preparation of figures.
We thank Dr. Szymon Godlewski and Dr. Mathias Ljungberg for helpful discussions. 
\end{acknowledgement}

\bibliography{Mendeley}

\end{document}